\documentclass[twocolumn,showpacs,preprintnumbers,amsmath,amssymb,prl]{revtex4}

\usepackage{dcolumn}
\usepackage{bm}
\usepackage{hyperref} 
\usepackage{amsmath}
\usepackage{amsfonts}     
\usepackage{graphics}  
\usepackage{psfrag} 
\usepackage{graphicx}    
\usepackage{epsfig}    
\usepackage{rotating}  
 \usepackage[latin1]{inputenc}
 \usepackage{color}
 \usepackage{rotating}
\usepackage{pdfpages}

\definecolor{DarkBlue}{rgb}{0.0,0.0,0.5}
\definecolor{DarkGreen}{rgb}{0.0,0.7,0.0}
\definecolor{LightBlue}{rgb}{0.5,1.0,1.0}
\definecolor{orange}{rgb}{1.0,0.4,0.0}
\definecolor{DarkYellow}{rgb}{0.7,0.7,0}
\definecolor{grey}{rgb}{0.5,0.5,0.5}

\begin{document}


\title{Unidirectional mechanical amplification as a design principle for an active microphone}

\author{Tobias Reichenbach}
\author{A.~J.~Hudspeth}

\affiliation{Howard Hughes Medical Institute and Laboratory of Sensory Neuroscience, The Rockefeller University, 1230 York Avenue, New York, NY 10065-6399}

\date{\today}

\begin{abstract}
Amplification underlies the operation of many biological and engineering systems.   Simple electrical, optical, and mechanical  amplifiers are reciprocal: the backward coupling of the output to the input equals the forward coupling of the input to the output. Unidirectional amplifiers that occur often in electrical and optical systems are special non-reciprocal devices in which the output does not couple back to the input even though the forward coupling persists. Here we propose  a scheme for unidirectional mechanical amplification that we utilize to construct an active microphone. We show that amplification  improves the microphone's threshold for detecting weak signals and that unidirectionality prevents distortion.
\end{abstract}


\maketitle

\begin{figure}[b]
\begin{center}
\includegraphics[width=88mm]{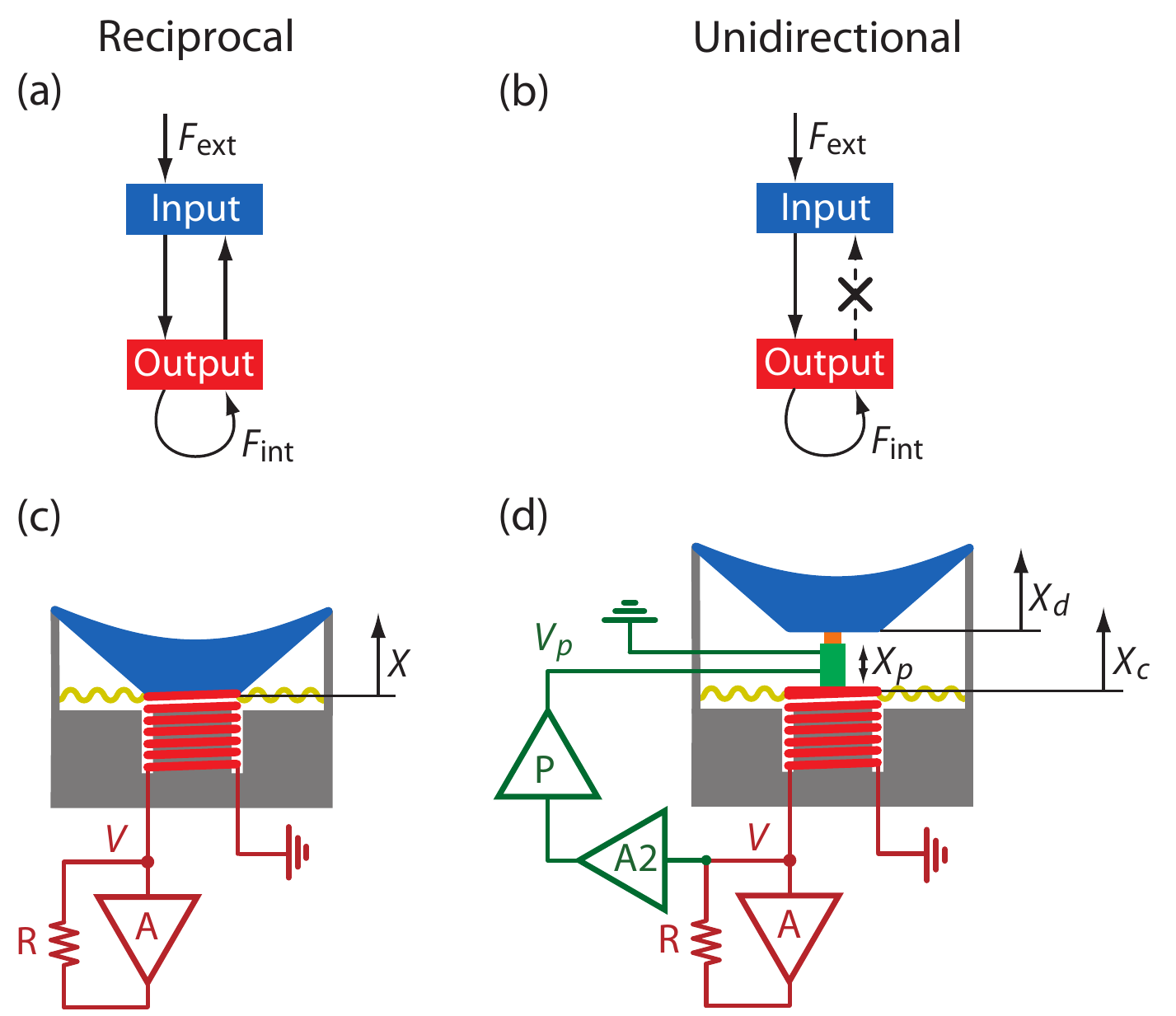}
\caption{\label{fig:mics} 
Amplifier and microphone designs. (a),(b), Schematic diagrams of reciprocal and unidirectional amplifiers.   (c),  In a dynamic microphone the spider (yellow) and diaphragm (blue) hold the voice coil (red) in a magnetic field. The coil's displacement can be amplified through amplifying the coil voltage $V$ (amplifier $A$) and feeding it back into the coil through a resistance $R$. (d), Unidirectionality results from a piezoelectric element (green) in series with an elastic element (orange)  placed between the coil  and diaphragm. The piezoelectric element's length change $X_p$  is controlled by a voltage $V_p$ that results from the coil voltage $V$ and that is adjusted through an amplifier $A2$ and a phase shifter $P$.  }
\end{center}
\end{figure}

Maxwell's reciprocity theorem in mechanics states that a force applied at  a point  A, such as an amplifier's input, produces a  displacement
 at a point B, say the amplifier's output, that equals the displacement at A caused by an equal force at B~\cite{potton-2004-67,Barber} [Fig.~\ref{fig:mics}(a)]. Similar reciprocal relations govern electrical circuits~\cite{Bleaney} and optics~\cite{potton-2004-67}. They represent a potential problem in  the design of amplifiers, for an ideal amplifier should operate unidirectionally: although the signal at the input  must control the output, to avoid feedback and distortion the output should not couple back to the input  [Fig.~\ref{fig:mics}(b)]. 
Reciprocity can be violated in electrical circuits through the use of semiconductors and in optical systems through Faraday rotation~\cite{potton-2004-67,Pozar}.  Amplifiers in electrical engineering and microwave devices therefore employ these effects to achieve unidirectionality~\cite{mccullen-2007-98,Pozar}.

Mechanical amplification can enhance the detection of a weak signal by raising its amplitude above the noise level. Biology employs this strategy in hearing for mechanosensitive hair cells in the vertebrate inner ear  actively amplify weak sounds and thereby greatly lower the threshold of hearing~\cite{robles-2001-81,hudspeth-2008-59}.  In contrast, microphones -- the ear's technological analogues -- are  passive devices that do not employ mechanical amplification but rely on subsequent electronic signal processing. One difficulty in implementing mechanical amplification in microphones is the reciprocity described above, which leads to undesired feedback and hence highlights the need for a mechanism of unidirectional mechanical amplification.  We have recently described how such a mechanism may operate in the mammalian inner ear, where hair cells provide piezoelectric coupling~\cite{ashmore-2008-88} between the input and output that can foster unidirectionality~\cite{reichenbach-2010,reichenbach-2010-105}. Here we show that a similar strategy can be used to implement unidirectional mechanical amplification in a microphone.

A dynamic microphone functions by the same principle as a dynamic speaker~\cite{Eagle}: a diaphragm is attached to a coil that moves in a magnetic field [Fig.~\ref{fig:mics}(c)]. Sound vibrates the diaphragm and thereby causes oscillations of the coil that electromagnetically induce a voltage. Such a system  serves as a speaker when an oscillatory electrical signal is applied to the coil; the consequent Lorentz force  vibrates the coil and hence the diaphragm, emitting sound.  The system's dual function as microphone and speaker therefore results from two potential  forces. An external sound force $F_\text{ext}$ acts on the diaphragm, representing the microphone's input, and the internal Lorentz force $F_\text{int}=IlB$ acts on the coil, corresponding to the output.  Here $I$ denotes the current in the coil, $l$ the coil's length, and $B$ the magnetic field.
To linear order the diaphragm's displacement $X$ then obeys
\begin{equation}
m\partial_t^2X+\lambda\partial_t X+KX=F_\text{int}+F_\text{ext},\label{eq:X}
\end{equation}
in which $m$, $\lambda$, and $K$ denote respectively the mass, damping coefficient, and stiffness of the coil together with the diaphragm and the entrained air. The voltage $V$ induced by coil motion is
\begin{equation}
V=Bl\partial_t X.\label{eq:V}
\end{equation}

Because of the system's dual  function as both speaker and microphone it can readily be converted into a mechanically active microphone [Fig.~\ref{fig:mics}(c)]. Indeed, positive feedback of the electrical signal in the coil can employ the internal Lorentz force to increase the coil's displacement resulting from an external sound force. To this end we have constructed a circuit in which the electrical signal in the coil is amplified  and fed back into the coil through a resistance $R$ [Fig.~\ref{fig:mics}(c)]. The electrical amplification is unidirectional with an adjustable gain $G$ achieved through an operational amplifier in the non-inverting feedback configuration (Supplementary material~\cite{reichenbach-2010-EPAPS}). The amplifier's output voltage $V_A$ in response to the input $V$ is then $V_A=GV$ and fullfills $V_A-V=IR$; these relations may be employed to express the voltage $V$ in terms of the current $I$:
\begin{equation}
V=\left(\frac{R}{G-1}\right)I.\label{eq:V_I}
\end{equation}
To analyze Equations~(\ref{eq:X}),~(\ref{eq:V}), and (\ref{eq:V_I}), we consider a pure tone of frequency $f$ that produces an oscillatory external force $F_\text{ext}=\tilde{F}_\text{ext}e^{i\omega t} + c.c.$ in which the tilde connotes the Fourier component, $\omega=2\pi f$, and and `$c.c.$' denotes the complex conjugate. Because the equations are linear, the resulting displacement $X$, voltage $V$, and current $I$ oscillate at the same frequency $f$: $X=\tilde{X}e^{i\omega t}+c.c.$, $V=\tilde{V}e^{i\omega t}+c.c.$, and $I=\tilde{I}e^{i\omega t}+c.c.$ The amplitudes follow as
\begin{equation}
\tilde{X}=\frac{1}{i\omega Z}\tilde{F}_\text{ext},~~\tilde{V}=\frac{Bl}{Z}\tilde{F}_\text{ext},~~\tilde{I}=\frac{(G-1)Bl}{RZ}\tilde{F}_\text{ext}\label{eq:X_act}
\end{equation} \
with the impedance
\begin{equation}
Z=i(m\omega-K/\omega)+ \lambda-l^2B^2(G-1)/R. \label{eq:Z_eff}
\end{equation}

Equation~(\ref{eq:Z_eff}) defines a resonant frequency $\omega_0=2\pi f_0=\sqrt{K/m}$ at which the imaginary part of the  impedance vanishes. The real part results from the damping $\lambda$ counteracted by the positive feedback. At a critical value of the gain $G_c=1+\lambda R/(l^2B^2)$, the real part of the impedance vanishes; this value therefore defines a transition from damped to undamped oscillation~\cite{Strogatz,Wiggins}. Nonlinearities control the system's behavior at this bifurcation and  yield a transition from damped to stable limit-cycle oscillations consistent with a supercritical Hopf bifurcation (Supplementary material~\cite{reichenbach-2010-EPAPS}).

  \begin{figure}[b]
\begin{center}
\includegraphics[width=87mm]{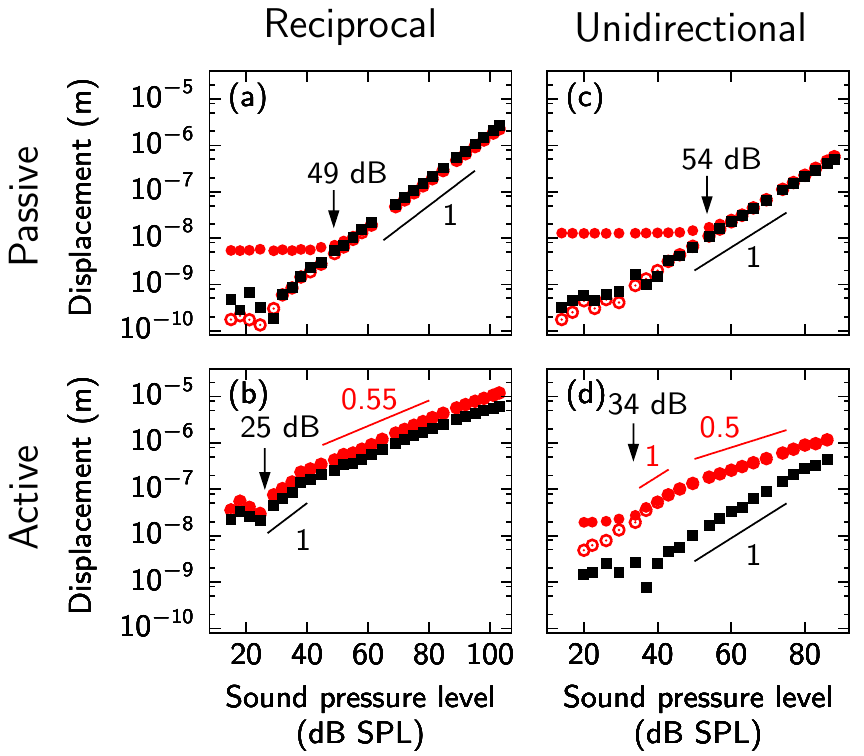}
\caption{\label{fig:response} (color online).
Linear and nonlinear responses.  The diaphragm's displacement (black squares) and the coil's displacement (red or grey circles) are presented as functions of the sound-pressure level for the reciprocal passive (a), the reciprocal active (b), the unidirectional passive (c), and the unidirectional active microphone (d) stimulated at the resonant frequencies of their active versions.
The diaphragm's displacement is obtained from interferometric measurements for which we report the Fourier component. The coil's displacement is inferred from the coil's voltage; we display the Fourier component (red or grey open circles) as well as the root-mean-square value multiplied by $\sqrt{2}$ (red or grey full circles). Arrows indicate the detection thresholds at which the response enters the noise floor.
}
\end{center}
\end{figure}
Operation at the Hopf bifurcation enhances the detection of sounds at the resonant frequency~\cite{eguiluz-2000-84,camalet-2000-97,PhysRevLett.91.128101,mccullen-2007-98}.  The active microphone's response is sharply tuned to the resonant frequency $f_0$ (Supplementary material~\cite{reichenbach-2010-EPAPS}) at which  the linear part of the response vanishes [Equation~(\ref{eq:Z_eff})] and the response becomes nonlinear. In our active microphone the response follows a power-law relation to stimulus amplitude with an exponent of about $0.5$  (Fig.~\ref{fig:response}(b), Supplementary material~\cite{reichenbach-2010-EPAPS}).   Faint signals cause a significantly larger response than in the passive microphone [Fig.~\ref{fig:response}(a),(b)]. Amplification acts only on the noise component at the resonant frequency and not on the remaining noise spectrum and thus improves the  signal-to-noise ratio and lowers the threshold for signal detection at the resonance. Because an active microphone constructed in this way detects only a small band of frequencies near its resonance, an ensemble of such systems tuned to different resonant frequencies is required to cover a broader frequency range. Such an array of active oscillators indeed underlies hearing in the vertebrate inner ear~\cite{robles-2001-81,hudspeth-2008-59}.

The active microphone described above operates reciprocally: the motion of the diaphragm, representing the input, is amplified by the coil's vibration, which corresponds to  the output. Unidirectionality can be achieved by placing a piezoelectric element, controlled by the coil's electrical signal, between the diaphragm and coil [Fig.~\ref{fig:mics}(d)]. The element's length change $\tilde{X}_p$ is then proportional to the coil displacement $\tilde{X}_c$: $\tilde{X}_p=-\alpha\tilde{X}_c$ with a complex coefficient $\alpha$ determined by the electrical circuit. Because the external sound force acts on the diaphragm and the internal Lorentz force on the coil, the diaphragm's displacement $\tilde{X}_d$ and the coil's displacement $\tilde{X}_c$  follow from 
\begin{equation}
i\omega A \begin{pmatrix} \tilde{X}_d\\ \tilde{X}_c \end{pmatrix}= \begin{pmatrix} \tilde{F}_\text{ext} \\ \tilde{F}_\text{int} \end{pmatrix}
\end{equation}
with the matrix
\begin{equation}
A=\begin{pmatrix}  Z_d +Z &  -(1+\alpha)Z \\  -Z & Z_c+(1+\alpha)Z  \end{pmatrix}.
\end{equation}
Here $Z_d$ denotes the impedance of the diaphragm, $Z_c$ the impedance of the coil, and $Z$ the impedance of a coupling element positioned between the piezoelectric element and the diaphragm [Fig.~\ref{fig:mics}(d)]. 

The piezoelectrical coupling breaks reciprocity: when $\alpha\neq 0$, the coupling of the coil to the diaphragm, given  by the matrix element $A_{12}$, differs from the coupling of the diaphragm to the coil, represented by $A_{21}$. At a critical value $\alpha_*=-1$  the coupling becomes unidirectional: the matrix element $A_{12}$ vanishes and with it the coupling from the coil to the diaphragm, whereas $A_{21}$ and thus the coupling from the diaphragm to the coil remains nonzero. Setting the coefficient $\alpha$ to its critical value $\alpha_*$ requires adjustment of both its amplitude and phase, which can be achieved through amplifying and subsequently phase-shifting the coil voltage (Fig.~\ref{fig:mics}(d) and Supplementary material~\cite{reichenbach-2010-EPAPS}). At the critical value $\alpha_*$ the displacements are
\begin{align}
\tilde{X}_d&=\frac{1}{i\omega(Z_d+Z)}\tilde{F}_\text{ext},  \label{eq:Xd} \\
\tilde{X}_c&=\frac{Z}{i\omega Z_c(Z_d+Z)}\tilde{F}_\text{ext} + \frac{1}{i\omega Z_c}\tilde{F}_\text{int}.  \label{eq:Xc}
\end{align}
Unidirectional coupling is manifest in these equations because the coil is displaced both by the external sound force and by the internal Lorentz force [Equation~(\ref{eq:Xc})], whereas only the sound force acts on the  diaphragm [Equation~(\ref{eq:Xd})].

Because amplification in the unidirectional active microphone acts through the internal Lorentz force  on the coil but not on the diaphragm, the coil's displacement exhibits a Hopf bifurcation at a critical gain whereas the motion of the diaphragm does not (Supplementary material~\cite{reichenbach-2010-EPAPS}). The system's reponse at the Hopf bifurcation reflects this difference:  the coil's displacement and voltage exhibit nonlinear behavior because their linear responses vanish at the bifurcation [Fig.~\ref{fig:response}(d)]. The diaphragm's displacement, however, shows a linear response for it does not encounter the amplification. 

Amplification in the unidirectional active microphone lowers the threshold for detecting signals at the resonant frequency. Indeed, a passive unidirectional microphone that omits amplification through elimination of the active microphone's positive-feedback circuit exhibits a considerably higher threshold [Fig.~\ref{fig:response}(c),(d)]. 
 Because the piezoelectrical coupling introduces additional noise, the signal-detection thresholds in the  passive and active unidirectional microphones  are slightly increased compared to their reciprocal analogues (Fig.~\ref{fig:response}).

\begin{figure}[t]
\begin{center}
\includegraphics[width=87mm]{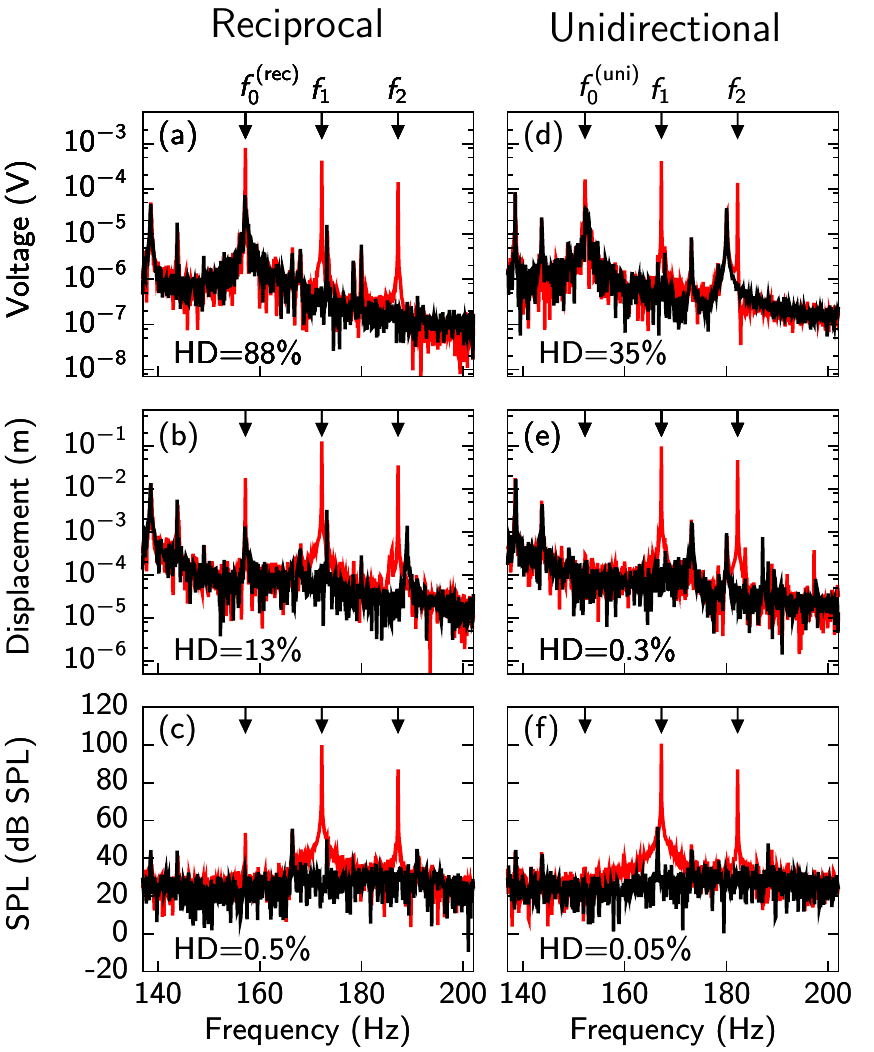}
\caption{\label{fig:dist} (color online).
Distortion products.  We report the Fourier spectrum of the coil's voltage (a, d), of the diaphragm's displacement (b, e), and of the sound-pressure level detected by an external microphone (c, f) for the reciprocal and the unidirectional active microphone. The system's reponse to stimulation at two frequencies $f_1$ and $f_2$ with $f_0=2f_1-f_2$ is shown in red or grey, and the background signal  without stimulation in black. The amount of distortion can be quantified through the harmonic distortion (HD, Supplementary material~\cite{reichenbach-2010-EPAPS}). 
}
\end{center}
\end{figure}
What is the benefit of unidirectionality in an active microphone? Amplification of the diaphragm requires energy that is spared with unidirectional coupling. The critical gain $G_c$ of the amplifier at which the Hopf bifurcation emerges is therefore lower in the unidirectional microphone than in the reciprocal one (Supplementary material~\cite{reichenbach-2010-EPAPS}).

Another important advantage is prevention of distortion. The nonlinear response of an active microphone near its Hopf bifurcation causes the formation of distortion products~\cite{julicher-2001-98}: when stimulated at two frequencies $f_1$ and $f_2$ the microphone also responds to linear combinations of these frequencies such as $2f_1-f_2$ and $2f_2-f_1$. A strong response results if one distortion product coincides with the microphone's resonant frequency $f_0$ [Fig.~\ref{fig:dist}(a)].  When amplification is reciprocal this distortion product is emitted from the active microphone because the coil transmits it to the diaphragm and a  sound results [Fig.~\ref{fig:dist}(a),(b),(c)]. Unidirectional amplification, in contrast, prevents the emission of such a distortion product. Although the distortion product appears in the coil's voltage and displacement owing to the coil's operation near a Hopf bifurcation, the distortion product is not transmitted to the diaphragm and is therefore not emitted as sound [Fig.~\ref{fig:dist}(d),(e),(f)].

An ensemble of distortion products results when amplification is reciprocal. The number and frequency of distortion products to which an active microphone reports are determined by the form of the nonlinearities that dominate near the bifurcation.  For example,  the normal form of the Hopf bifurcation induces only the cubic distortion products $2f_1-f_2$ and $2f_2-f_1$  in response to stimulation at $f_1$ and $f_2$. However, as described above, detection of signals across a certain frequency range requires an array of active microphones with distinct resonances. Presentation of $f_1$ and $f_2$ to such an array results in a cascade of combination tones at frequences $f_1\pm n(f_2-f_1),~n\in \mathbb{N}$, for the distortion products at $2f_1-f_2$ and $2f_2-f_1$  are emitted by microphones tuned to these frequencies and interact with the stimuli at frequencies $f_1$ and $f_2$ as well as themselves to create  other distortion products such as $3f_1-f_2$, $3f_2-f_1$, and so on.  Such cascades of distortion products have been recorded from the high-frequency region of the mammalian inner ear, where amplification is reciprocal~\cite{robles-1997-77}. 

Unidirectional coupling prevents the cascade of distortion products. Although distortion products such as $2f_1-f_2$ and $2f_2-f_1$ are formed in the coil, they are not emitted and hence cannot create further distortion products. The resulting  reduction in the number of distortion products represents  a significant advantage of the  unidirectional active microphone over its reciprocal counterpart. Theoretical considerations indicate that the mammalian ear may employ the same strategy for detection of frequencies below about $2$~kHz~\cite{reichenbach-2010}, which are  employed predominantly in  human speech~\cite{Baken} and music~\cite{Olson}.
 
 In summary, we have described a scheme for unidirectional mechanical amplification as well as its implementation in an active microphone that is ultra-sensitive and non-distorting. This technology has  potential applications in contexts that require the sensitive detection of mechanical signals at specific frequencies, for example sonar, sonography, and perhaps the detection of seismic  or even gravitational waves.



\onecolumngrid

\newpage

\setcounter{page}{1}

\begin{center}

{\large \bf
Unidirectional mechanical amplification as a design principle\\for an active microphone
 }\\

\vspace*{0.5cm}

{\large Tobias Reichenbach and A. J. Hudspeth}\\

\vspace*{0.5cm}

{\large \bf Supplementary EPAPS Document}

\end{center}

\vspace*{1.5cm}

{\bf \large Construction details}\\

\begin{figure*}[t]
\begin{center}
\hspace*{-0.0cm}\includegraphics[width=160mm]{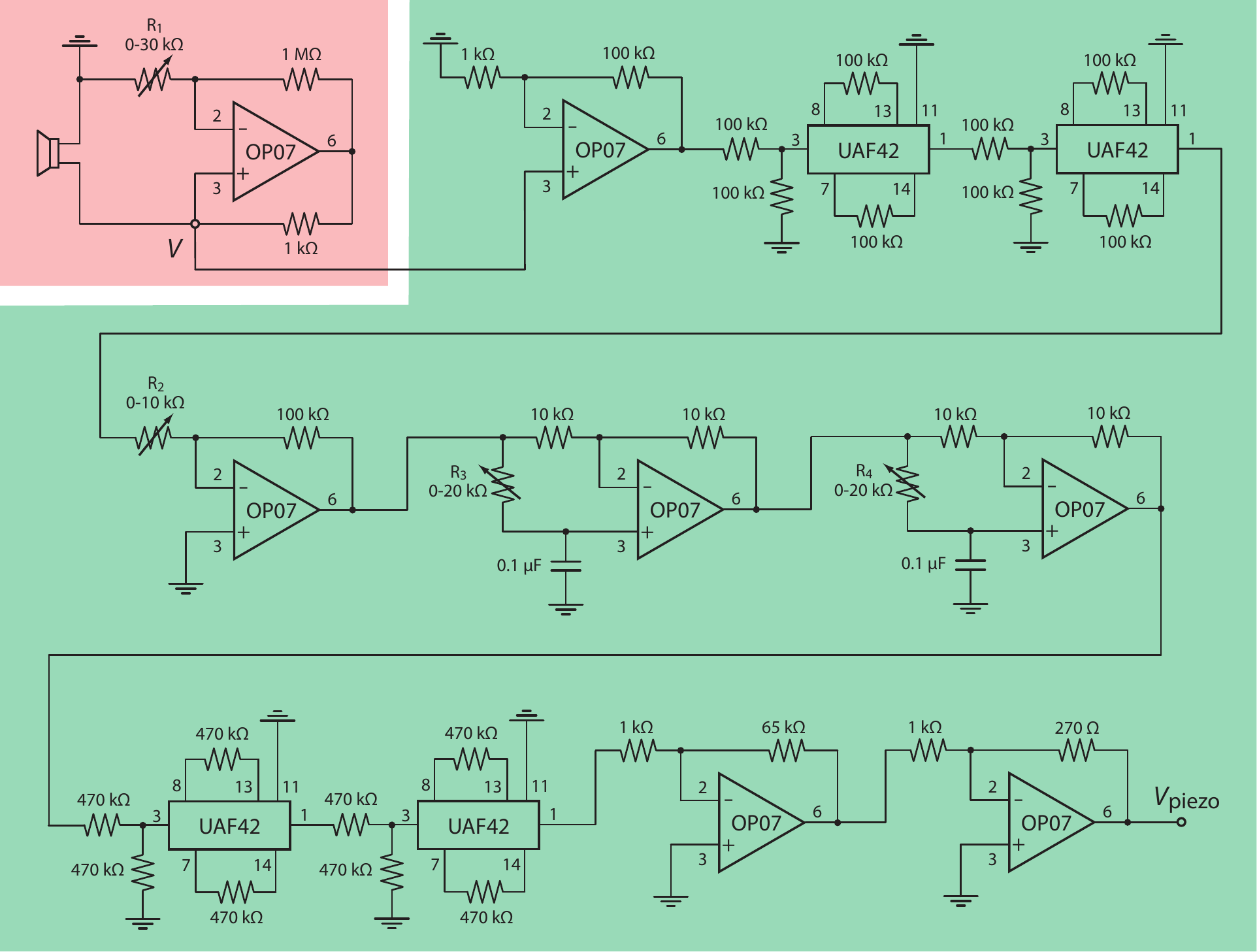}
\caption{\label{fig:circuit}
 Electrical circuit. The electrical circuit indicated in red provides positive feedback for mechanical amplification in the microphone and constitutes the circuit indicated in red in Fig.~1(d),(e). Its feedback gain can be adjusted by varying the resistance $R_1$.  The circuitry indicated in green converts the coil voltage $V$ into a voltage $V_\text{piezo}$ that is fed into an amplifier of maximal voltage $150$~V and gain $10\times$ to yield the piezo voltage $V_p$.  The gain of this circuit can be controlled through the variable resistance $R_2$ and its phase through the variable resistances $R_3$ and $R_4$. Four two-pole low-pass filters are constructed with the help of filters (UAF42; Texas Instruments). This arrangement  constitutes the circuit indicated in green in Fig.~1(e).}
\end{center}
\end{figure*}

\begin{figure}[t]
\begin{center}
 \includegraphics[scale=1]{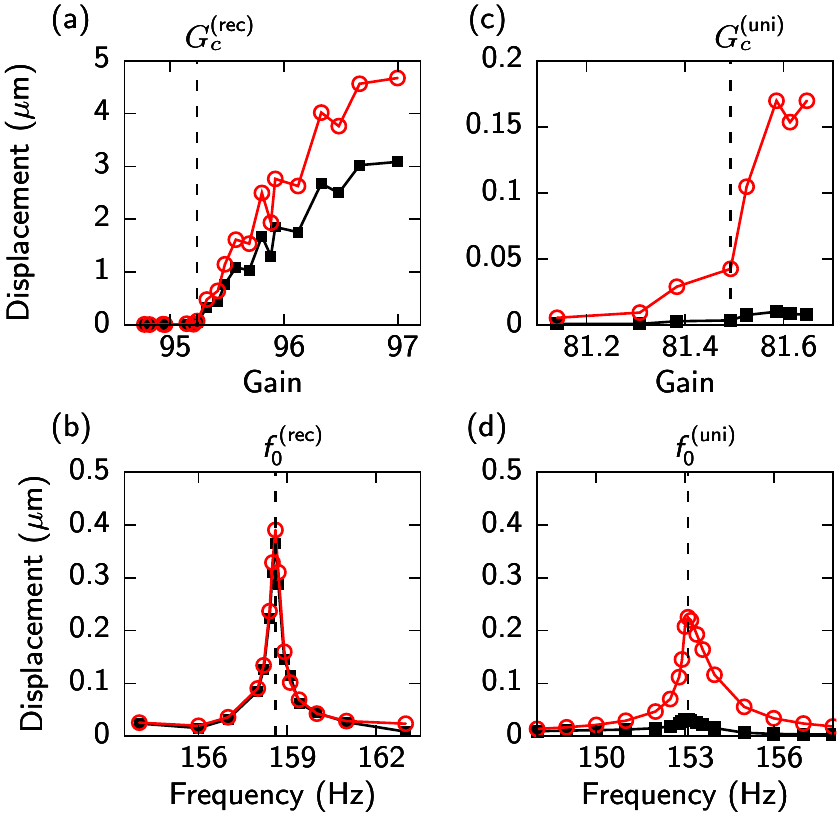} 
\caption{\label{fig:hopf}
Hopf bifurcation. We show the displacement of the coil (red circles) and of the diaphragm (black squares) for the reciprocal  active (a,b) and the unidirectional  active microphone (c,d).
(a),(c), The reciprocal active and the unidirectional active microphone exhibit supercritical Hopf bifurcations at the critical values $G_c^\text{(rec)}\approx 95.2$ and $G_c^\text{(uni)}\approx81.5$ of the gain in amplifier $A$.
In the absence of an external stimulus, and apart from noise, the system is quiescent below $G_c$. For gains exceeding  $G_c$
spontaneous limit-cycle oscillations emerge in the coil. The diaphragm displacement reflects the Hopf bifurcation in the reciprocal but not in the unidirectional system.
(b),(d), When operating at the Hopf bifurcation and stimulated by sound of different frequencies at a constant sound-pressure level of $60$~dB SPL,
the  responses of the reciprocal and the unidirectional active microphones are tuned sharply around their resonant frequencies of $f_0^\text{(rec)}\approx158.6$~Hz and $f_0^\text{(uni)}\approx153.1$~Hz. The diaphragm's motion approximately equals the coil's motion in the reciprocal microphone, but is much smaller than the coil's displacement in the unidirectional system.
}
\end{center}
\end{figure}

\begin{figure}[t]
\begin{center}
 \includegraphics[scale=1]{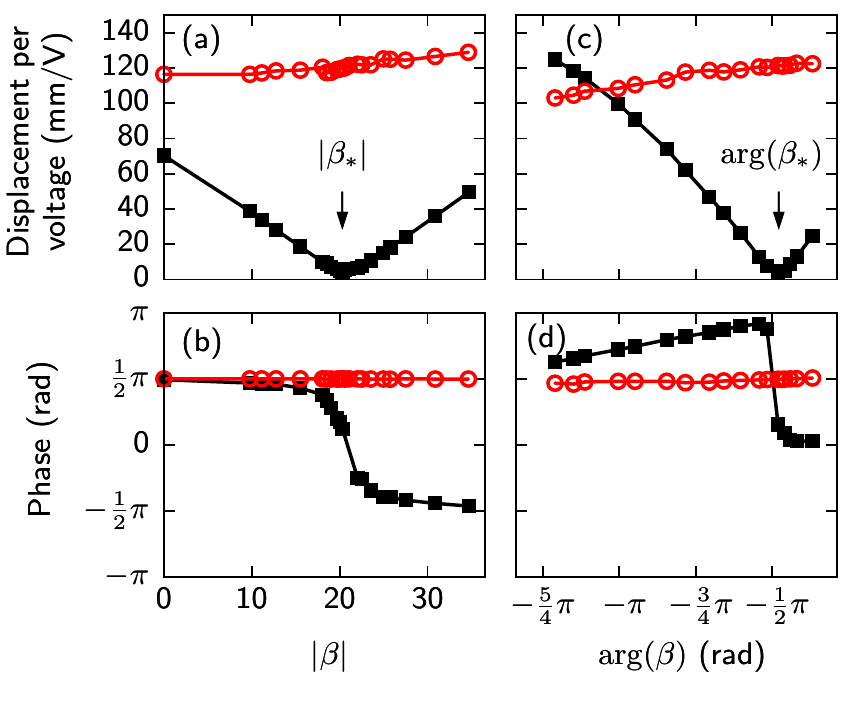} 
\caption{\label{fig:uni}
Adjusting unidirectionality.  Unidirectionality emerges when the ratio $\beta$ of the piezoelectric-element voltage to the coil voltage, $\beta\equiv\tilde{V_p}/\tilde{V}$, which relates linearly to the coefficient $\alpha$, is tuned to  the critical value $\beta_*$ of the correct amplitude and phase.   We show  the  diaphragm displacement per coil voltage, $\tilde{X_d}/\tilde{V}$ (black squares) as well as the coil displacement per coil voltage, $\tilde{X_c}/\tilde{V}$ (red circles)  that result when the coil but not the diaphragm is stimulated.  
(a),(b), The amplitude and phase of the displacement per voltage when the magnitude of $\beta$ is varied and its phase is held constant at the critical value $\arg(\beta_*)$. (c),(d), The amplitude and phase  of the displacement per voltage when the phase of $\beta$ is varied and its magnitude held constant at the critical value $|\beta_*|$. The coil's displacement per voltage is largely independent of $\beta$ (a,b,c,d). However, the diaphragm's displacement almost vanishes at $\beta_*$, showing that the forces on the coil are not transmitted to the diaphragm and thus demonstrating unidirectionality (a,c).  The diaphragm's displacement does not vanish completely at $\beta_*$ because of nonlinearities in the piezoelectric element and because of noise. In agreement with theoretical analysis, the diaphragm's displacement undergoes a phase change of $\pi$ around $\beta_*$  (b,d).
}
\end{center}
\end{figure}

We modified a $6.5$-inch-diameter  speaker (Soundstream) by detaching the diaphragm from the voice coil and inserting between them a piezoelectric element  $10$~mm in length and capable of a displacement of $9.1~\mu$m at a maximal voltage of $150$~V (Thorlabs). The speaker's spider was modified to yield a smaller coil displacement $X_c$ per volt  in the range of the piezoelectric element. We connected the system to an electrical circuit that provided electrical feedback to the voice coil and hence mechanical amplification (Fig.~\ref{fig:circuit}).  A variable resistance controlled the gain of this circuit and allowed adjustement to a Hopf bifurcation (Fig.~\ref{fig:hopf}). Another part of the circuit was designed to allow adjustment of the piezoelectric element's voltage to different magnitudes and phases (Fig.~\ref{fig:circuit}). Because the voltage in the electrical circuit operated at $\pm15$~V, we sent the signal into an amplifier with a fixed gain of $10\times$ and with a maximal output voltage of $150~$V; its output was connected to the piezoelectric element. To counteract high-frequency noise in the circuit we added four two-pole low-pass filters (Fig.~\ref{fig:circuit}).  To adjust unidirectionality (Fig.~\ref{fig:uni}) we drove the voice coil with two additional piezoelectrical elements  that were anchored in the speaker's wall and attached to the coil. Sounds were generated with Mathematica (Wolfram Research) and presented through a $4$-inch-diameter speaker (Boss Audio Systems). For distortion products two frequencies $f_1$ and $f_2$ were generated independently and presented through two distinct speakers. The harmonic distortion (HD) of the signal was calculated as
\begin{equation}
\text{HD}=\sqrt{\frac{V_{2f_1-f_2}^2}{V_{2f_1-f_2}^2+V^2_{f_1}+V^2_{f_2}}}
\end{equation}
in which $V_{f}$ denotes the voltage at frequency $f$.\\

{\bf \large Nonlinearities}\\

This section concerns the nonlinear behavior of the active microphones. We focus our discussion on the reciproal active microphone; the unidirectional  case follows similarly.

A system operating near a Hopf bifurcation can be described by the normal form~[15]
\begin{equation}
\partial_t z=(a+i\omega_0)z+(b+ic)|z|^2z + o (z^3).
\label{eq:normal_form}
\end{equation}
Here $z$ is a complex variable that, in the case of the active microphone, can encode the coil's displacement $X$ and velocity $\partial_t X$. The resonant frequency is given by $\omega_0$ and $a,b,c$ are real coefficients with $a=0$ representing the Hopf bifurcation.  The linear part of Equation~(\ref{eq:normal_form}) can be derived from Equations (1-3). The nonlinear part results from nonlinear forces in the system. Although those forces may have different quadratic and cubic (as well as higher-order) contributions in the variables $X$ and $\partial_t X$, a  nonlinear transformation to the $z$-variable exists such that $z$ agrees with $X,\partial_t X$ to linear order, such that the quadratic nonlinearities disappear, and such that the cubic nonlinearities take the form $|z|^2z$ [Equation~(\ref{eq:normal_form})]. Higher-order nonlinearities can be cast into the $U(1)$-symmetric forms $|z|^4z,~|z|^6z$, and so on.  

When the normal form is  periodically forced by $\tilde{f}e^{i\omega_0 t}$ at the resonant frequency and at the Hopf bifurcation ($a=0$), and when the resulting displacement is small such that nonlinearities higher than the cubic can be ignored, the Fourier component $\tilde{z}$ at $\omega_0$ obeys
\begin{equation}
0=(b+ic)|\tilde{z}|^2\tilde{z}+\tilde{f}
\end{equation}
and thus exhibits a nonlinear response
\begin{equation}
|\tilde{z}| \sim |\tilde{f}|^{1/3}
\end{equation}
with a power-law exponent of $1/3$. 

Our recordings show an exponent of about $1/2$ (Fig.~3b,d) instead of the exponent $1/3$ that characterizes the response close to the Hopf bifurcation. How can this exponent be understood through the speaker's nonlinear behavior? One  nonlinear force in a speaker comes from the magnetic field generated by the moving voice coil that attracts the iron in the pole piece surrounding it~[11]. 
At the coil's resting position, $X=0$, this force is directed perpendicular to the coil's motion and balanced because of an approximately symmetric geometry. For displacements $X\neq 0$, however, a component $F_\text{mag}$ of this force emerges which acts to restore the coil to its resting position; for small $X$ the relative strength of this component to the net force is proportional to the coil's displacement $X$. The strength of the net force depends on the product of the magnetic field $H$ generated by the coil and the pole piece's resulting magnetization $M(H)$. Together we arrive at
\begin{equation}
F_\text{mag}\sim XHM(H).
\end{equation}
The magnetic field $H$ is proportional to the coil's velocity and thus, for sinusoidal motion, to the phase-shifted displacement $X$. 

The dependence of the magnetization $M$ on the magnetic field $H$ is more complicated. For small magnetic fields the two are proportional: $M\sim H$. Small displacements $X$ induce a small magnetic field $H$ and therefore a cubic nonlinear force: $F_\text{magn.}\sim X^3$.
Larger magnetic fields, however, lead to saturation at a maximal magnetization $M_\text{max}$ and therefore a  magnetization of $-M_\text{max}$ if the magnetic field is reversed. For these larger  field strengths the magnetization  may be approximated by a step function:
\begin{equation}
M(H)\approx\left \{\begin{array}{lcl} -M_\text{max} & \text{for} & H<0 \\
M_\text{max} & \text{for} & H>0
\end{array} \right. .
\end{equation}
For larger displacements $X$ the nonlinear force therefore assumes the approximate form
\begin{equation}
F_\text{mag}\sim |X|X.
\end{equation}
This nonlinear force accordingly makes a contribution of $|z|z$ in Equation~(\ref{eq:normal_form}):
\begin{equation}
\partial_t z=(a+i\omega_0)z+(d+ie)|z|z + o (z^2)
\label{eq:quadratic_form}
\end{equation}
with real coefficients $d,e$. In agreement with our observation, the term  $|z|z$  induces a response $|\tilde{z}|=|\tilde{f}|^{1/2}$ at the bifurcation and at the resonant frequency.
We hence argue that the exponent $1/2$ emerges in our active microphones because  we operate at relatively high sound-pressure levels that induce saturation of the induced magnetic field in the pole piece and thus elicit a combination of higher-order nonlinearities summing to a contribution of  $|z|z$ that dominates the cubic nonlinearity. If we could detect signals at the much lower sound-pressure levels at which noise dominates in our system, we would expect to find a regime in which the magnetization operates in the unsaturated regime and a cubic nonlinearity emerges. We note that other nonlinearities, such as that in the spider's stiffness~[11], 
 contribute as well. We expect, however,  that they contribute to cubic nonlinear behavior. We have observed a cubic nonlinear response in a smaller speaker ($4$-inch-diameter, Boss Audio Systems) for which the nonlinear contribution $F_\text{mag}$ described above is apparently not dominant.

\end{document}